\documentstyle[aps,epsf,prl,multicol]{revtex}

\begin{document}
\title{Creation of Skyrmions in a Spinor Bose-Einstein Condensate}

\author{Karl-Peter Marzlin}
\address{Fachbereich Physik der Universit\"at Konstanz, 
   Postfach 5560 M674, D-78457 Konstanz, Germany}

\author{Weiping Zhang and Barry C. Sanders}
\address{Department of Physics and Centre for Lasers and Applications, \\
Macquarie University, Sydney, New South Wales 2109, Australia}
\date{\today}
\maketitle
\begin{abstract}
We propose a scheme for the creation of 
skyrmions (coreless vortices) in a Bose-Einstein condensate with
hyperfine spin $F=1$.
In this scheme, four traveling--wave laser beams, with
Gaussian or Laguerre-Gaussian transverse profiles, 
induce Raman transitions with
an anomalous dependence on the laser polarization,
thereby generating the optical potential required for producing skyrmions.
\end{abstract}

$ $ \\
03.75.Fi, 32.80.-t
\begin{multicols}{2}

The recent experimental success in all--optical trapping of
an atomic Bose-Einstein condensate (BEC)\cite{ketterle98}
opens the prospect of studies into the internal structure of 
spinor BECs, including the possibility of creating vortex states
without core, or skyrmions, in the BEC\cite{ho98,machida98,machida99,ruostekoski99,yip99}.
Skyrmions, which do not have an ordinary vortex core due to the
spin degree of freedom, offer a myriad of new physical phenomena
beyond those presented by other vortex states.
One feature is the reduction of kinetic energy associated with the rotation
by transferring this energy to the spin. 
Another interesting property of skyrmions is that they do not represent
a topological excitation, although their flux is vortex-like far away
from the line of symmetry \cite{woelfle90}.  
In this paper, we propose an optical method to create skyrmions
in a condensate of alkali atoms.
This method, which employs laser beams and Raman transition to generate
skyrmions in the BEC, is related to proposals for the creation of
vortices in a single--component BEC\cite{marzlin97,bolda98}. 

For a BEC with hyperfine spin $F=1$,
a skyrmion wave function, which is axially symmetric
around the $z$-axis, has the form\cite{ho98}
\begin{equation} 
  \left ( \begin{array}{c} \psi_{-1} \\ \psi_0 \\  \psi_1 
  \end{array} \right ) = \sqrt{\rho} 
  \left ( \begin{array}{c} 
  \cos^2 \left (\frac{\beta}{2} \right ) \\  
  \sqrt{2} e^{i \phi} \cos \left (\frac{\beta}{2} \right )
  \sin \left (\frac{\beta}{2} \right ) \\  
  e^{2i\phi} \sin^2 \left (\frac{\beta}{2} \right ) 
  \end{array} \right ) \; ,
\label{skyrmion}
\end{equation} 
with $\rho(z,r_\perp)$ the total density of atoms,
${\bf r_\perp} = (x,y)$ the transverse coordinate vector
and $r_\perp = | {\bf r_\perp} |$ the radial distance from the $z$-axis.
The angle~$\phi$ corresponds to the orientation of~${\bf r_\perp}$
in the $x$-$y$--plane.
The function~$\beta(r_\perp)$,
which characterizes the spin state of the condensed atoms,
is related to the superfluid velocity ${\bf v}_s$ of the system by
\begin{equation} 
  {\bf v}_s = \frac{\hbar}{M r_\perp} [1-\cos(\beta(r_\perp))] 
  {\bf e}_\phi \; ,
\end{equation} 
where ${\bf e}_\phi \equiv (\cos\phi ,-\sin\phi,0)$ 
is the unit vector in the~$\phi$ direction.
Because a skyrmion has no vortex core,
$\beta(0)=0$ must hold in order to
avoid a singularity.
For $\beta(r_\perp)=\pi/2$, the superfluid velocity 
reduces to that of an ordinary vortex state.

To create a skyrmion we consider a BEC which first is magnetically trapped
in the $m=-1$ hyperfine state. We assume that the trap is then switched off and
the optical potential is applied to the BEC. Thus, our initial state is
given by $ \psi_{-1} = \sqrt{\rho(r_\perp,z)} $ and $\psi_0=\psi_1 =0$.
Our objective is to design an optical potential which transfers this state
into the state (\ref{skyrmion}). 

\begin{figure}[t]
\epsfxsize=8cm
\hspace{1cm}
\epsffile{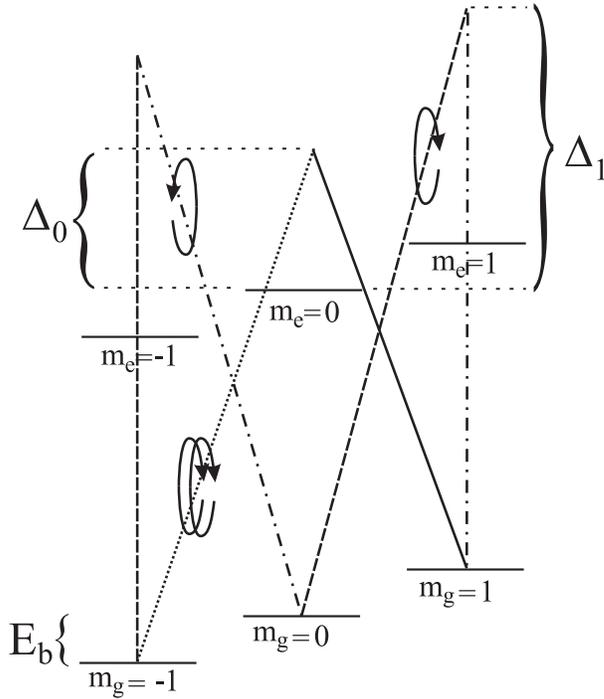}
\caption{\label{levels} A scheme of Raman transitions needed for the creation
of skyrmions. The solid line corresponds to laser beam 0, the dashed line to
beam 1, the dot-dashed line to beam -1, and the dotted line to beam 2. 
Diagonal transitions describe ordinary transitions with photons polarized 
in $x$-direction. Vertical transitions are anomalous transitions with photons
polarized in beam direction ($z$-axis).}
\end{figure}

Generally a coherent optical potential for atoms is created by applying
several highly detuned laser beams which induce Raman
transitions between different angular momentum states or internal states
of the atomic BEC (cf Fig.~\ref{levels}).
In order to preserve the cylindrical symmetry of the skyrmions along the 
$z$-axis, the laser beams must also propagate
along this axis.
This restriction raises the following challenge: the polarization orientation
of laser beams propagating along the $z$-axis is generally presumed to be
in a superposition of ${\bf e}_x$ and ${\bf e}_y$.
However, as a Raman transition
of such beams can only effect transfers of the type $\psi_{-1}$ to $\psi_1$,
the possibility of creating skyrmions of the form~(\ref{skyrmion}) seems to
be excluded.

Fortunately the flexibility in the transverse structure of laser beams permits
a portion of the laser mode to be polarized along the
axis of propagation.
Mathematically, this contribution to longitudinal polarization has its origin
in the vanishing divergence of the electric field for any laser beam;
that is,
\begin{equation} 
  \mbox{div} {\bf E}({\bf r}) =0 \; . 
\label{div0} \end{equation} 
For a laser beam propagating along the $z$-axis,
with transverse mode function
$u({\bf r_\perp})$ and polarization vector ${\bf e}_x$,
the divergence condition~(\ref{div0}) makes it necessary that the
complete positive-frequency part of the electric field is given by
\cite{bemerkung}
\begin{equation} 
  {\bf E}^{(+)}({\bf r}) = e^{\pm ikz} e^{-i \omega t}
   \left[ i {\bf e}_x u({\bf r_\perp}) \mp {\bf e}_z
    \frac{1}{k} u^\prime ({\bf r_\perp})  \right] ,
\label{Efield} \end{equation} 
where $u^\prime ({\bf r_\perp})$ denotes the derivative of $u$ with respect to $x$.
The second term on the right-hand side of Eq.\ (\ref{Efield})
is essential for inducing transitions to~$\psi_0$ component.


Our proposal includes four traveling--wave laser beams and a homogenous
magnetic field ${\bf B} = B {\bf e}_z$. The purpose of the
magnetic field is to lift the degeneracy of the ground states
$| m_g \rangle$ with $m_g=0,\pm 1$ being the magnetic quantum number
of the states. We write the corresponding Hamiltonian in the form
\begin{equation}
  H_B = E_B \sum_{m_g=-1}^1 m_g |m_g \rangle \langle m_g | \; .
\label{bham} \end{equation}  

The laser beams induce
transitions between the $F=1$ manifold of the atomic ground state
and an $F=1$ manifold of excited atomic states, such as the
$D_1$ line between $^2$S$_{1/2}(F=1)$ and  $^2$P$_{1/2}(F=1)$
in Rubidium ($^{87}$Rb). 
All four beams are linearly polarized in the $x$-direction
and have a large detuning to suppress spontaneous
emission. Each beam has a different mode structure:
we include a Gaussian beam with frequency $\omega_0$ and a 
Laguerre-Gaussian beam of second order with frequency $\omega_2$.
Both beams are copropagating along the positive $z$-axis. 
Our scheme further includes two copropagating Laguerre-Gaussian beams
(along the negative $z$-axis)
of first order with opposite orbital angular momentum and frequencies
 $\omega_1$ and  $\omega_{-1}$, respectively.
All beams are assumed to have
the same width $w$ (where~$w$ is the width parameter for the
Gaussian envelope of each beam). The electric field of
each beam then is of the form (\ref{Efield}),
whereby the transverse mode functions
are approximately given by
\begin{equation} 
  u_l = A_l \frac{r_\perp^l}{w^l}e^{il\phi}  e^{-r_\perp^2/w^2} 
  \quad , \quad l=-1,0,1,2\; ,
\label{modes}
\end{equation} 

The interaction of the atoms with the laser beams in the electric
dipole approximation is given by
\begin{equation} 
  H_{\mbox{{\scriptsize int}}} = \hbar \sum_{l=-1}^2
  \hat{\Omega}_l e^{-i\omega_l t} + \mbox{H.c.} \; .
\end{equation} 
where $\hat{\Omega}_l$ is the Rabi-frequency operator for the 
$l^{\rm th}$ laser beam, defined by
\begin{eqnarray} 
  \hbar \hat{\Omega}_le^{-i\omega_l t} &\equiv& 
   - {\bf d}^\dagger \cdot {\bf E}^{(+)}_l  \\
   &=&-\sum_{m_e,m_g} \langle m_e| {\bf d}^\dagger \cdot {\bf E}^{(+)}_l
      | m_g\rangle |m_e\rangle \langle m_g| \nonumber
\end{eqnarray} 
where $m_e$ denotes the magnetic quantum numbers for
the excited states. Using standard techniques
for Clebsch-Gordon coefficients (eg, see Ref.~\cite{ct90}), we find
for each laser beam the expression
\begin{eqnarray} 
   \hbar \hat{\Omega}_l  &=& -{\cal D} e^{\pm ikz}
	 \Big[ \frac{ i u_l}{2} ( |0_e \rangle \langle 1_g| +
	|-1_e \rangle \langle 0_g| 
  \nonumber \\ & &  \hspace{1.7cm} +
	|1_e \rangle \langle 0_g| + |0_e \rangle \langle -1_g| )
  \nonumber \\ & &
  \mp \frac{ u_l^\prime }{k\sqrt{2}} ( |1_e \rangle \langle 1_g| -
  |-1_e \rangle \langle -1_g| ) \Big]  
\end{eqnarray} 
with~${\cal D}$ is a reduced matrix element. For simplicity we assume
${\cal D}$ to be real. The first contribution
describes the usual transitions which are induced by $x$-polarized laser beams.
The term proportional to $u^\prime_l$ 
has its origin from the small $z$-polarized term which
results from the vanishing divergence of the electric field.

The optical potential which can be deduced from such a configuration
for largely detuned laser beams has the general form
(see, e.g., Ref.~\cite{marzlin98})
\begin{equation} 
   V_{\mbox{{\scriptsize opt}}} = \hbar \sum_{l,l^\prime =-1}^2
	e^{i(\omega_{l^\prime}-\omega_l)t} \frac{ \hat{\Omega}_{l^\prime}
	^\dagger \hat{\Omega}_l }{ 
	\omega_l -\omega_{\mbox{{\scriptsize res}}}  } \; ,
\end{equation} 
where $\omega_{\mbox{{\scriptsize res}}}$ is the resonance frequency
for the electronic transition. The total Hamiltonian for this system
then takes the form 
\begin{equation} 
  H =   \int d^3x \sum_{m_g,m_g^\prime } 
  \Psi_{m_g}^\dagger \left \langle
  m_g \left |  H_{\mbox{{\scriptsize 1-p}}} 
   \right | m_g^\prime \right \rangle \Psi_{m_g^\prime} 
  + H_{\mbox{{\scriptsize nl}}}
\end{equation} 
with the one-particle contribution
\begin{equation} 
  H_{\mbox{{\scriptsize 1-p}}} = 
  \frac{{\bf p}^2}{2M} + H_B + V_{\mbox{{\scriptsize opt}}}
\end{equation} 
and the nonlinear interaction energy \cite{law98}
\begin{eqnarray} 
  H_{\mbox{{\scriptsize nl}}} &=& \frac{\lambda_s}{2} 
  \sum_{m_g,m_g^\prime} \int d^3x \Psi_{m_g}^\dagger 
  \Psi_{m_g^\prime }^\dagger \Psi_{m_g^\prime} \Psi_{m_g}
  \nonumber \\ & & 
  + \frac{\lambda_a}{2} \int \Big[ (\Psi_1^\dagger)^2 (\Psi_1)^2
  +  (\Psi_{-1}^\dagger)^2 (\Psi_{-1})^2 
    \nonumber \\ & & \hspace{1cm}
  + 2 \Psi_{1}^\dagger \Psi_{0}^\dagger\Psi_{1}\Psi_{0}
  + 2 \Psi_{-1}^\dagger \Psi_{0}^\dagger\Psi_{-1}\Psi_{0}
    \nonumber \\ & &  \hspace{1cm}
  - 2 \Psi_{1}^\dagger \Psi_{-1}^\dagger\Psi_{-1}\Psi_{1}
  + 2 (\Psi_{0}^\dagger)^2\Psi_{1}\Psi_{-1}
   \nonumber \\ & &  \hspace{1cm}
  + 2 \Psi_{1}^\dagger \Psi_{-1}^\dagger (\Psi_{0})^2 \Big] d^3x
\end{eqnarray} 
where $\lambda_i \equiv 4 \pi \hbar^2 a_i/M$ is proportional to the corresponding
scattering length $a_i$.
Transforming to the interaction picture, with respect to $H_B$ of
Eq.~(\ref{bham}), the optical potential assumes the form
\begin{eqnarray} 
  V_{\mbox{{\scriptsize opt}}} &=&  \frac{{\cal D}^2}{\hbar} 
	\sum_{l,l^\prime =-1}^2
	\frac{e^{i(\omega_{l^\prime}-\omega_l)t}  }{ 
	\omega_l -\omega_{\mbox{{\scriptsize res}}}  } \Big \{
  ({\bf 1} - |0\rangle\langle 0|) \frac{ u_{l^\prime}^{\prime *}
   u_l^\prime }{2k^2}
   \nonumber \\ & & 
   + ({\bf 1} + |0\rangle\langle 0| 
     +e^{2iE_bt}|1\rangle\langle -1|
   \nonumber \\ & &  \hspace{15mm}  
     +e^{-2iE_bt}|-1\rangle\langle 1|) 
    \frac{ u_{l^\prime}^{*} u_l }{4}
   \nonumber \\ & &  
   \pm i (e^{-iE_bt}|0\rangle\langle 1| - e^{iE_bt}|0\rangle\langle -1|) 
    \frac{ u_{l^\prime}^{*} u_l^\prime}{2\sqrt{2}k}
   \nonumber \\ & &  
   \mp i (e^{iE_bt}|1\rangle\langle 0| - e^{-iE_bt}|-1\rangle\langle 0|) 
    \frac{ u_{l^\prime}^{\prime *} u_l}{2\sqrt{2}k} \Big \}
\label{vopt2} \end{eqnarray} 

It is now our task to derive from this general expression a special
potential that is suitable for the creation of skyrmions.
As our initial state is the ground-state for a BEC trapped in $m_g=-1$,
it becomes obvious from Eq.~(\ref{skyrmion}) that we require
matrix elements of the form $|1\rangle\langle -1| e^{2i\phi}$ ,
$|0\rangle\langle -1| e^{i\phi}$ and so on.
As the transverse mode function $u_l$ is proportional to $e^{il\phi}$,
its derivative $u_l^\prime$ is approximately proportional to~$e^{i(l-1)\phi}$. This relation permits the following construction.

The detunings $\Delta_l \equiv \omega_l - \omega_{\mbox{{\scriptsize res}}}$
of the laser beams are the most significant frequencies in the problem.
To construct the desired matrix elements we assume that $\Delta_2
\approx \Delta_0$ and $\Delta_{-1} \approx \Delta_1$. However, $\Delta_0$
and $\Delta_1$ should be sufficiently different so that a Raman
transition involving the beams 0 and 1 is suppressed.
Assuming that the second largest frequency is given by the
magnetic interaction energy $E_B$
allows the rotating-wave approximation to be done
with respect to this frequency in Eq.~(\ref{vopt2}).
By assuming that
\begin{eqnarray}  
  \delta\omega_{02} &\equiv&  \omega_0 -\omega_2 + 2E_B \ll E_B, \; 
                          \omega_0-\omega_2 ,	\\
  \delta\omega_{11} &\equiv&  \omega_1 -\omega_{-1} -E_B \ll E_B, \; 
                          \omega_1-\omega_{-1}	,
\end{eqnarray} 
many terms can be neglected, and we arrive at
\begin{eqnarray} 
  V_{\mbox{{\scriptsize opt}}} &\approx&  
    \frac{ {\cal D}^2  }{ 2\hbar k^2  } 
    ({\bf 1} - |0\rangle\langle 0|) \left \{ 
    \frac{|u_0^\prime|^2 +|u_2^\prime|^2 }{\Delta_0} +
    \frac{|u_1^\prime|^2 +|u_{-1}^\prime|^2 }{\Delta_1} \right \}
  \nonumber \\ & & 
    + \frac{ {\cal D}^2 }{4\hbar} 
    ({\bf 1} + |0\rangle\langle 0|) \left \{ 
    \frac{|u_0|^2 +|u_2|^2 }{\Delta_0} +
    \frac{|u_1|^2 +|u_{-1}|^2 }{\Delta_1} \right \}
  \nonumber \\ & & 
    + \frac{ {\cal D}^2}{4\hbar \Delta_0} 
    \{ |1\rangle\langle -1| u_0^* u_2 e^{i\delta\omega_{02}t} + \mbox{ H.c} \}
  \nonumber \\ & & 
    - \frac{i{\cal D}^2}{2\sqrt{2} \hbar k \Delta_1} 
      \{ |0\rangle\langle 1|  u_1^* u_{-1}^\prime e^{i\delta\omega_{11}t}
          - \mbox{ H.c} \}
  \nonumber \\ & & 
    + \frac{i {\cal D}^2  }{2\sqrt{2} \hbar k \Delta_1}
      \{ |0\rangle\langle -1| u_{-1}^* u_1^\prime  e^{-i\delta\omega_{11}t}
         - \mbox{ H.c} \} .
\label{vopt3} \end{eqnarray} 

This term can be further simplified because the derivative
of $u_l$ typically scales as $1/w$ compared to $u_l$ itself. Including 
the prefactor $1/k$ that appears together with each derivative, we see that
every term with a derivative is suppressed by a factor of
\begin{equation} 
\label{varepsilon}
  \varepsilon \equiv \frac{1}{kw} = \frac{\lambda}{2\pi w} \; .
\end{equation} 
As the width of a laser beam always exceeds its wavelength
$\lambda$, the factor~(\ref{varepsilon}) is always much smaller than one.
For instance, for an optical transition ($\lambda = 795$ nm) a quite strongly
focused beam with a width of 5 $\mu$m has $\varepsilon \approx 0.025$.
This allows us to develop $ V_{\mbox{{\scriptsize opt}}}$ in a power series
in $ \varepsilon$. Before doing so we observe that in Eq.~(\ref{vopt3})
the transitions between $|0\rangle$ and $|\pm 1\rangle$ are suppressed by
a factor of $\varepsilon$ as compared to the transition between
$|1\rangle$ and $|-1\rangle$. Since this is undesirable for the production of
skyrmions we assume that the intensity $A_0$ (and hence $u_0$)
of the Gaussian laser beam defined by Eq.~(\ref{modes}) 
will be of the order of $\varepsilon$ as compared
to the other prefactors $A_l$. Inserting the mode functions (\ref{modes})
into Eq.~(\ref{vopt3}) and developing $ V_{\mbox{{\scriptsize opt}}}$ to 
first order in $\varepsilon$, we find
\begin{eqnarray} 
    V_{\mbox{{\scriptsize opt}}} &\approx&  
    \frac{ {\cal D}^2 e^{-2\bar{r}^2}}{4\hbar} \Big\{ 
    ({\bf 1} + |0\rangle\langle 0|) \left ( 
    \frac{A_2^2}{\Delta_0} \bar{r}^4 +
    \frac{2 A_1^2 }{\Delta_1} \bar{r}^2 \right )
  \nonumber \\ & & 
    + \left ( |1\rangle\langle -1| e^{2i\phi} \bar{r}^2
        \frac{A_0 A_2}{\Delta_0}
        e^{i\delta\omega_{02}t}
    + \mbox{ H.c} \right )
  \nonumber \\ & & 
    +\frac{i\sqrt{2} A_1^2}{\Delta_1}\varepsilon  \bar{r} \Big [
    |0\rangle\langle 1| e^{-i\phi} (1- \bar{r}^2(1+e^{-2i\phi}))
    e^{i\delta\omega_{11}t}
  \nonumber \\ & & 
    +  |0\rangle\langle -1| e^{i\phi} (1- \bar{r}^2(1+e^{2i\phi}))
    e^{-i\delta\omega_{11}t}
    - \mbox{ H.c} \Big ] \Big \}
\label{vresult} \end{eqnarray} 
Eq.~(\ref{vresult}) is the main result of this paper. It consists
of several parts which will now be analyzed.

The first row in Eq.~(\ref{vresult}) represents an optical trapping
potential if the laser detuning $\Delta_l$ is positive. For $\bar{r} 
\equiv r_\perp/w \ll 1$ it is harmonic. 
Its physical origin is a Raman transition which
returns to the same internal hyperfine level. As the state $|m_g=0\rangle$
has two possibilities ($|m_e=\pm1\rangle$) to make such a transition
the trapping potential is twice as strong as compared two the states
$|m_g=\pm1\rangle$. Note also that this term is of order $\varepsilon^0$
and therefore represents the dominant contribution.

The second row describes ordinary Raman transitions between the hyperfine
states $|m_g=-1\rangle$ and $|m_g=1\rangle$. As laser 2 corresponds
to a Laguerre-Gaussian beam of second order, an orbital angular momentum 
of $2\hbar$ will be transferred to the atoms, as it is required for
the formation of the skyrmion (\ref{skyrmion}).
We remark that the same result could have been
achieved by replacing the Gaussian beam 0 by a weak Laguerre-Gaussian beam
of order 1, and beam 2 by a Laguerre-Gaussian beam of order $-1$. 
The optical potential would only be changed in that the trapping potential
in the first row of Eq.~(\ref{vresult})
would become harmonic (apart from the overall exponential 
factor). From an experimental point of view this alternative may be
advantageous because it only requires two different types of laser beams 
(Laguerre-Gaussian of order 1 and -1) instead of four types.

The third and the fourth row in Eq.~(\ref{vresult}) describe 
anomalous Raman transitions in which the atoms absorb or emit
a photon with polarization ${\bf e}_z$ in one of the two processes involved.
In Fig.~\ref{levels} this corresponds to a vertical transition from
$|m_g=1\rangle$ (or $|m_g=-1\rangle$)  to 
$|m_e=1\rangle$ (or $|m_e=-1\rangle$). The possibility of these transitions
is a direct consequence of the full mode structure (\ref{Efield}).
Since these vertical transitions are proportional to the derivative
$u_l^\prime $ of the transverse mode functions, no orbital angular momentum
is transferred even though the corresponding beams are of Laguerre-Gaussian
type $\pm 1$ (more precisely, for $\bar{r} \ll 1$ the terms proportional to
$\exp [\pm 2 i \phi]$ in Eq.~(\ref{vresult}) are negligible). 
The second process which completes the Raman transition
is an ordinary transition from $|m_g\rangle$ to $|m_e=m_g\pm 1\rangle$.
Since the corresponding laser beam is of Laguerre-Gaussian type $\pm 1$
the total orbital angular momentum transferred to the atoms is $\pm \hbar$.
As a result, the anomalous Raman transitions create simple vortex states
in $|m_g=0\rangle$ and doubly excited vortices in $|m_g=1\rangle$.

It might be counterintuitive that the $z$-polarized contribution to
the electric field (\ref{Efield}), which usually can safely be neglected,
does now play an important role for the creation of skyrmions. In fact,
its contribution remains as small as usual compared to the dominant
parts (diagonal elements of $V_{\mbox{{\scriptsize opt}}}$). 
The important point in our proposal is that we can make all
matrix elements inducing internal transitions equally large
by using a weak laser beam 0 so that the usual Raman transition
$m_g=-1 \leftrightarrow m_g=1$ is not stronger than the anomalous 
Raman transitions.

As the diagonal elements of $V_{\mbox{{\scriptsize opt}}}$ are
much larger than the off-diagonal elements they provide a strong trapping
potential. According to Eq.~(\ref{skyrmion}) a skyrmion is a superposition
of the ground-state in $m_g=-1$  and a singly (doubly) excited vortex 
state in $m_g=0 \; (m_g=1)$, respectively. Because of the strong trapping
potential these states all have different energies. 
Thus, to induce transitions using the
off-diagonal elements of  $V_{\mbox{{\scriptsize opt}}}$ these
matrix elements need to rotate at some frequency to match a resonance 
condition. This can be done by adjusting the frequencies $\delta 
\omega_{02}$ and $\delta \omega_{11}$ appropriately.
The efficiency of the skyrmion production strongly depends on
$\delta\omega_{02}$ and $\delta\omega_{11}$. 

We have numerically examined the time evolution of a spin-1 BEC in
the potential (\ref{vresult}). Our simulations are based on the
split-step method \cite{splitstep} in two spatial dimensions
for a three-component nonlinear Schr\"odinger equation
(It is assumed that the BEC is homogeneously extended  along the $z$-axis
with a length of 50 $\mu$m). 
We consider a BEC of 15000 Rubidium atoms 
($M=1.45\times 10^{-25}$ kg,
 $ a_s = 5.4$ nm, $a_a = -0.05$ nm \cite{ho98}).
For a much larger number of atoms 
our method seems to be unsuitable because the potential (\ref{vresult})
has a maximum height due to the exponential envelope. A strong repulsive
interaction between the atoms then pushes the atoms out of the trapping
area.

\begin{figure}[t]
\epsfxsize=8cm
\hspace{1cm}
\epsffile{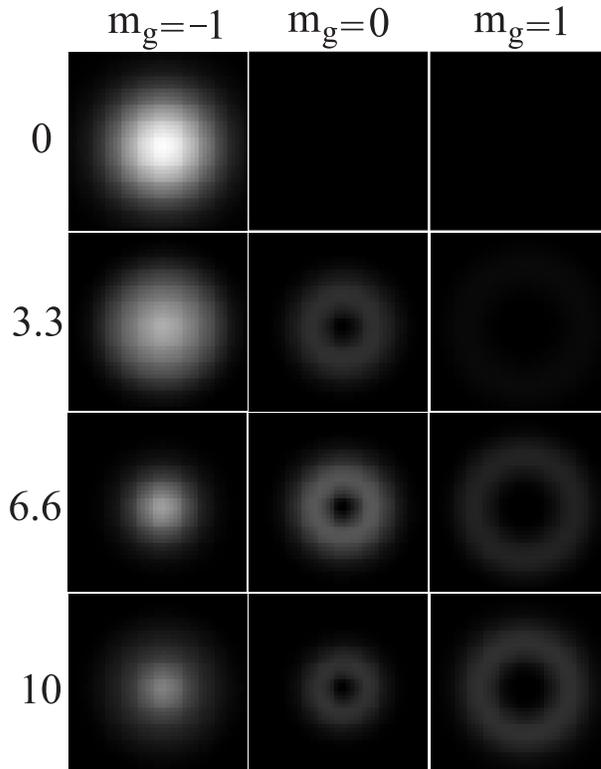}
\caption{\label{numresults} Time evolution of a three-component BEC
under the influence of the potential (\ref{vresult}). Shown is the 
spatial density of the three hyperfine states $m_g =0,\pm1$ at four
instants of time (given in ms). The initial state is the ground state
(of a harmonic trap) in the state $m_g=-1$.}
\end{figure}

For the optical potential (\ref{vresult}) we consider the situation that
$A_2^2/\Delta_0 = 2 A_1^2/\Delta_1$ holds and the amplitude of the Gaussian
beam is given by $A_0 = 3 \varepsilon A_2/\sqrt{2}$. The effective Rabi
frequency of the trapping potential is then given by
${\cal D}^2 A_2^2/(4\hbar^2 \Delta_0)$ which we take to be 66 kHz
(For a detuning $\Delta_0$ of 1 GHz this corresponds to a laser intensity
of about 25 mW). The width of the laser beams is 5 $\mu$m. 
For the results shown in
Fig.~\ref{numresults} we have used $\delta\omega_{02} = -2.4$ kHz and
$\delta\omega_{11} = 2.95$ kHz. At time $t=6.6$ ms this results in 
a superposition of internal states with about 27\% of the atoms
in state $m_g=-1$, about 34\% in the state $m_g=0$, and 39\% in state
$m_g=1$. For various other choices of $\delta\omega_{02}$ and 
$\delta\omega_{11}$ we have found as little as 4\% population in the initial
state $m_g=-1$ and as much as 70\% in one of the other two states.
The phase of the three components agrees very well with
the situation described by the state (\ref{skyrmion}). Animations
of the time evolution are available on the internet \cite{animations}. 

{\bf Acknowledgement:} K.-P.~M. would like to thank the Optik Zentrum
Konstanz for financial support.
This project has been supported by an Australian Research Council Small Grant.


\end{multicols}

\begin{thebibliography}{99}
\bibitem{bemerkung} This can also be derived by taking the appropriate
	limit (essentially $k^2 -\kappa^2 \approx k^2$) in Eqs.\ (3.3) and
	(4.1) of Ref.\ \cite{barnett94}.
\end{thebibliography}
\end{document}